\documentclass[aps,prl,showpacs,twocolumn]{revtex4}
\usepackage{graphicx}
\input{epsf}

\begin{document}
\title {Macroscopic quantum tunneling in globally coupled series arrays of Josephson junctions}

\author{M. V. Fistul}

\affiliation {Theoretische Physik III, Ruhr-Universit\"at Bochum,
D-44801 Bochum, Germany}
\date{\today}
\begin{abstract}
We present a quantitative analysis of an escape rate for switching
from the superconducting state to a resistive one in series arrays
of globally coupled Josephson junctions. A global coupling is
provided by an external shunting impedance. Such an impedance can
strongly suppress both the crossover temperature from the thermal
fluctuation to quantum regimes, and the macroscopic quantum
tunneling (MQT) in short Josephson junction  series arrays.
However, in large series arrays we obtain an enhancement of the
crossover temperature, and a giant increase of the MQT escape
rate. The effect is explained by excitation of a {\it
spatial-temporal charge instanton} distributed over a whole
structure. The model gives a possible explanation of recently
published experimental results on an enhancement of the MQT in
single crystals of high-$T_c$ superconductors.
\end{abstract}


\maketitle
Great attention has been devoted to an experimental and
theoretical study of dc biased series arrays of Josephson
junctions \cite{Likh,Barbara,Nymeier,Haviland}. Such a system
displays diverse fascinated  nonlinear classical and macroscopic
quantum-mechanical phenomena. E.g. a resistive state of Josephson
junction series arrays can show synchronized behavior
\cite{Likh,Barbara}, and this effect has been used in Josephson
voltage standard devices \cite{Nymeier}. As we turn to a region of
small dc bias currents and low temperatures, the macroscopic
quantum-mechanical phenomena start to play a role. Thus, a quantum
phase superconductor-insulator transition has been observed in
artificially prepared series arrays of a small size
$Al/Al_20_3/Al$ junctions \cite{Haviland}. All these effects
strongly depend on the interaction between Josephson junctions.

This field of research, i.e. the macroscopic quantum phenomena in
spatially extended superconducting systems, has been boosted even
further by recent discovery of macroscopic quantum tunneling (MQT)
in single crystals of layered high-$T_c$ superconductors
\cite{HTSC-QM1,HTSC-QM2}. At low temperatures the MQT determines
the escape rate of the switching from the superconducting state to
a resistive one. Although the MQT of a single Josephson phase has
been found long time ago in low-$T_c$ lumped Nb Josephson
junctions \cite{Tinkham,Clarke},  MQT in layered high-$T_c$
superconductors has shown many novel features. A most unexpected
result is that the MQT escape rate $\Gamma_{MQT}$ in high-$T_c$
superconductors is {\it four order of magnitude} larger than the
MQT escape rate for a lumped Josephson junction having the same
parameters \cite{HTSC-QM2}. Moreover, the crossover temperature
$T^{*}$ from the thermal fluctuation regime to the MQT regime
increases in respect to a lumped Josephson junction. In these
experiments it was also found that the escape rate $\Gamma_T$ in
the thermal fluctuation regime did not differ from the escape rate
of a single Josephson junction.

Layered high-$T_c$ superconductors can be modelled as a stack (a
series array) of intrinsic Josephson junctions \cite{Kleiner}.
A modern fabrication technique allows to prepare single
crystals of layered high-$T_c$ superconductors with an extremely
homogeneous distribution of critical currents of intrinsic
Josephson junctions, and a low level of dissipation
\cite{HTSC-QM1,HTSC-QM2}. Since in the model of independent
Josephson junctions the escape rate $\Gamma$ is just
proportional to $N$, an enhancement of the MQT observed in layered
high-$T_c$ superconductors stems from an interaction between
intrinsic Josephson junctions. All experimental observations
receive a natural explanation in a simple model \cite{Tachicki} of
Josephson junctions series array with an intrinsic charge
interaction between nearest-neighbor Josephson junctions
\cite{Matchida,Fistul-CoulInt}. Moreover, such a model allows
quantitative comparison with experimental results, and a good
agreement has been found as the Debye screening length is of the
order of superconducting layer thickness \cite{Fistul-CoulInt}.

However, the authors of Ref. \cite{HTSC-QM2} proposed an other
model in order to explain a giant increase of the MQT escape rate.
In this model intrinsic Josephson junctions are {\it globally
coupled} due to the presence of an external shunting impedance $Z$
(see schematic in Fig. 1).
An influence of electromagnetic environment, and in particular, a
shunting impedance $Z$ on the MQT in a lumped Josephson junction
has been studied long time ago in Refs. \cite{EDM,Ingold,Legett}.
It was shown that the presence of a small shunting impedance can
lead to a strong {\it suppression} of the MQT in a lumped
Josephson junction. Therefore, natural questions arise: what is a
role of shunting impedance $Z$ in the macroscopic quantum dynamics
of large Josephson junction series arrays ($N>>1$) and what is a
most appropriate model in order to explain a giant increase of the
MQT escape rate in Josephson junction series arrays?

In order to answer these questions we carry out a quantitative
analysis of the escape rate $\Gamma(I)$ in globally coupled
Josephson junction series arrays. A global coupling is provided by
an external shunting impedance $Z$ (see Fig. 1). We obtain, and it
is a main result of the paper, that if the electromagnetic
environment strongly suppresses the MQT in a lumped Josephson
junction, the role of an external shunting impedance is diminished
in large Josephson junction arrays, and  the standard
quantum-mechanical behavior is recovered. Therefore, the globally
coupled Josephson junction series arrays can show a giant increase
of the MQT escape rate with a strong dependence on a number of
junctions. Moreover, the MQT escape rate is tunable in a wide
region by a simple change of $Z$. Such tuning of the MQT escape
rate can be very promising for a modern field of quantum
information processing \cite{Qcomp}.

\begin{figure}
\includegraphics[width=2in,angle=-90]{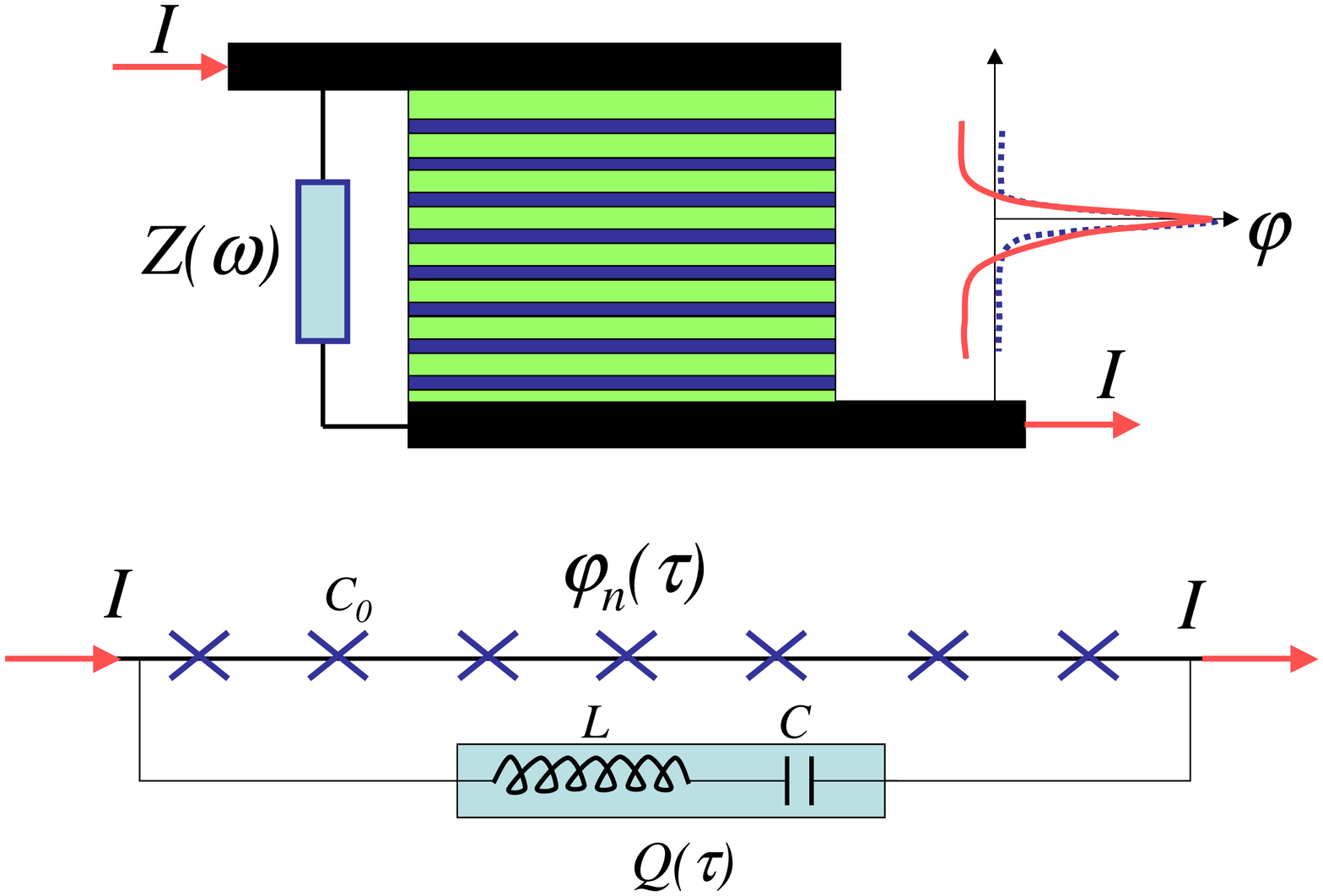}
\caption{Schematic of a dc biased layered high-$T_c$
superconductor and a series array of Josephson junctions. A
strongly localized instanton (dashed line) and a charge instanton
with long tails (solid lines) are shown.} \label{Schematic}
\end{figure}

A Josephson junction series array is characterized by the set of
Josephson phases $\varphi_n (\tau)$, where number $n$ changes from
$1$ to $N$. Moreover, electrodynamics of a shunting impedance $Z$
is described by a flowing charge $Q(\tau)$. We will consider a
particular case as all resistive effects are small, i.e. the
shunting impedance has only a reactive part, and it contains
inductor $L$ and capacitor $C$ in series. In order to obtain the
escape rate $\Gamma$ we use "instanton technique"
\cite{Coleman,Affleck,Ingold}, and therefore, $\varphi_n (\tau)$
and $Q(\tau)$ are periodic functions of the imaginary time $\tau$
varying from $0$ to $\hbar/(k_B T)$ ($T$ is the temperature). The
escape rate is determined by the action $S$ as
\begin{eqnarray} \label{escape-action}
\Gamma~\simeq~\int Dq(\tau) D\varphi_n(\tau) \exp
\left[-\frac{S\{q,\varphi_n \}}{\hbar} \right]~,\nonumber \\
S=\int_0^{\hbar/k_B T} L(\tau) d\tau ~~~~~~~~~~~~
\end{eqnarray}
and the Lagrangian of a series array with the shunting impedance
is written as
\begin{eqnarray} \label{Lagr}
L~=~\sum_n \frac{1}{2\omega_p^2}[ \dot \varphi_n (\tau)]^2+
\frac{1}{2\omega_R^2}[ \dot q(\tau)]^2+\frac{1}{2}q(\tau)^2+\nonumber \\
+\sum_n U_n+i\frac{\alpha}{\omega_R} \sum_n q(\tau)\dot \varphi_n
(\tau)~,~~~~~~~~~~~~~~~\nonumber \\
U_n(\varphi)=\cos \varphi_n (\tau)+j \varphi_n
(\tau)~,~~~~~~~~~~~~
\end{eqnarray}
where $ j=I/I_c$ is the normalized external dc current, and $I_c$
is the nominal value of the critical current of a single junction.
Here, $\omega_p$ is the plasma frequency of a single Josephson
junction in the absence of dc bias. The Lagrangian is expressed in
units of $E_J$, where $E_J$ is the Josephson energy of a single
junction. The $q=Q/\sqrt{E_JC}$ is the normalized charge flowing
through the impedance $Z$. The shunting impedance is characterized
by the resonance frequency $\omega_R=1/\sqrt{LC}$, where $L$ and
$C$ are the impedance inductance and capacitance, accordingly. The
coupling between the Josephson junction series array and the
shunting impedance branch is described by parameter
$\alpha=\sqrt{L_J/L}$, and $L_J=2e/(\hbar I_c)$ is the Josephson
inductance.

Integrating (\ref{escape-action}) over $q(\tau)$ \cite{Feynman} we
obtain the effective action $S_{eff}$ that depends on the
variables $\varphi_n(\tau)$ only
\begin{eqnarray} \label{Action-eff}
S_{eff}\{\varphi_n \}=\sum_n \int_0^{\frac{\hbar}{k_B T}} d\tau
\frac{1}{2\omega_p^2}[ \dot \varphi_n (\tau)]^2+U_n~~~~~~~~~~~~\nonumber \\
+\frac{\alpha^2}{2}\int_0^{\frac{\hbar}{k_B
T}}\int_0^{\frac{\hbar}{k_B T}} d \tau_1 d \tau_2 \cdot ~~~~~~~~~~~~~~~~~ \nonumber \\
\cdot G_T(\tau_1-\tau_2)[\sum_n \dot \varphi_n (\tau_1)][\sum_m
\dot \varphi_n (\tau_2)]~,~~~~
\end{eqnarray}
where the kernel  $G_T(\tau)$ is determined as
\begin{eqnarray} \label{Kernel}
G_T(\tau)=\frac{k_BT}{\hbar}\sum_j \frac{ e^{i\omega_m
\tau}}{\omega_m^2+\omega_R^2},~~~~~~~~~\nonumber \\
\omega_m=m(2\pi k_BT)/\hbar,~ m=\pm 1,\pm 2....
\end{eqnarray}
Thus, the last term in Eq. (\ref{Action-eff}) presents an
effective global charge interaction, that is due to current
fluctuations flowing through an external shunting impedance.

In the escape experiments $E_J>>\hbar \omega_p$, and the switching
to a resistive state occurs as the dc current $I$ is close to
$I_c$, and therefore $(j-1) \ll 1$. In this case  the potential
$U_n(\varphi)$ is written as
\begin{equation} \label{Lagr2}
U_n(\varphi)~=~(1-j) \varphi_n (\tau)-\frac{\varphi_n^3
(\tau)}{6}~.
\end{equation}
The escape rate is determined by the particular solution
$\varphi_n(\tau)$ providing the extremum of effective action
(\ref{Action-eff}). At high temperatures such a solution is
determined by extremum points of the potential $U_n$, and it is
written as
$$ \varphi_n^T=2\sqrt{2(1-j)}\delta_{nl}-\sqrt{2(1-j)}~.
$$
Here, $l$ is a junction number where the fluctuation occurs. Since
this solution does not depend on the time $\tau$, we can
immediately conclude that the last term in (\ref{Action-eff}) does
not give contribution to the escape rate exponent
$\Gamma_T~\simeq~\exp(-S\{\varphi_n^T \}/\hbar $. Note here that
an  absence of the dependence of the escape rate exponent in the
thermal fluctuation regime on a number of junctions $N$ is a
generic property of Josephson junctions series arrays with a
charge interaction \cite{Fistul-CoulInt}.

However, the crossover temperature from the thermal fluctuation
regime to the MQT can be strongly enhanced by such a global
coupling. Indeed, using the method elaborated in
\cite{Ingold,Coleman,Affleck} we obtain that at high temperatures
the optimal fluctuation $\varphi_n(\tau)$ around an extremum point
has a form:
\begin{equation} \label{OptflHT}
\varphi_n(\tau)~=~e^{\frac{2\pi i k_B T \tau}{\hbar}} \phi_n~,
\end{equation}
where the eigenfunctions $\phi_n$ are the solution of the nonlocal
and inhomogeneous equation:
\begin{eqnarray} \label{EigenfunctHT}
\xi^2\phi_n+\frac{2\alpha^2\xi^2
\omega_p^2}{\xi^2+\omega_R^2}\sum_n \phi_n -2\omega_0^2
\delta_{ln}\phi_n ~=~(\lambda-\omega_0^2) \phi_n~, \nonumber \\
\xi=\frac{2\pi k_B T}{\hbar}.~~~~~~~~~~~~~~~~~~~~~~~~~~~~~~
\end{eqnarray}
Here, $\lambda$ are the eigenvalues of the Eq.
(\ref{EigenfunctHT}), $\omega_0~=~\omega_p[2(1-j)]^{1/4}$ is the
dc bias dependent frequency of oscillations on the bottom of
potential well, $U_n(\varphi)$.  The crossover temperature $T^*$
is determined by the condition that there is the eigenvalue
$\lambda=0$ \cite{Affleck,Ingold}. In a global coupling case (Eq.
(\ref{EigenfunctHT})) the crossover temperature is obtained as a
solution of the particular transcendent equation:
\begin{equation} \label{Transeq}
\xi^{*4}+\frac{2N\alpha^2\omega_p^2
\xi^{*2}}{\xi^{*2}+\omega_R^2}[\xi^{*2}-(1-2/N)\omega_0^2]=\omega_0^4~,
\end{equation}
where $\xi^*=\frac{2\pi k_B T^*}{\hbar}$. Thus, one can see that
the crossover temperature $T^*$ is strongly suppressed
for short arrays ($N~\simeq~1$) for both cases, namely, "inductive"
($\omega_R \ll
\omega_0$, and $\alpha~\simeq~1$) or "capacitive" ($\omega_R \gg \omega_0$, and $\alpha~>>~1$)
types of an external impedance.
However,  $T^*$ recovers
to the value $T^*=\hbar \omega_0/(2\pi k_B)$ for long arrays
($N>>1$). Typical dependencies  of $T^*(N)$ are shown in Fig.
2.
\begin{figure}
\includegraphics[width=3in,angle=0]{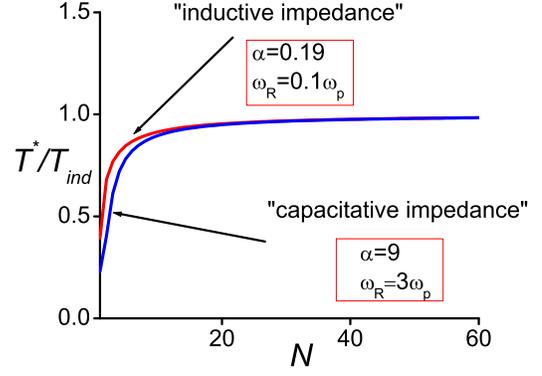}
\caption{Typical dependencies of the crossover temperature
$T^*(N)$ on a number of junctions $N$. Both cases of inductive (upper curve) and
capacitative (lower curve) impedance for particular sets of parameters are shown.}\label{CrossTdep}
\end{figure}

Now we turn to the MQT regime, where the extremum point of the
action $S_{eff}\{ \varphi_n\}$ is the "tau-dependent" instanton
(bounce) solution. At zero temperature and in the presence of an
external impedance $Z$, a spatial-temporal instanton solution
satisfies the equation:
\begin{equation} \label{Instantoneq}
\frac{1}{\omega_p^2}\ddot{\varphi}_n(\tau)+ \alpha^2 \sum_m \int_0^\infty
d \tau_1G_0(\tau-\tau_1)
\ddot{\varphi}_m-\frac{dU_n}{d\varphi}~=~0~,
\end{equation}
where $G_0(\tau)=\frac{1}{2\omega_R}\exp(-\omega_R |\tau|)$. The
solution of Eq.(\ref{Instantoneq}) has a following form: a large
bounce solution localized on a particular junction $l$,
$\varphi_l(\tau)=f(\tau)$, and a small spatial-temporal tail
solution distributed over a whole array (see schematic in Fig. 1,
solid line). The Fourier transform of the instanton tail is
obtained as
\begin{equation} \label{Tailsolution}
\sum_{n \neq l} \varphi_n(\omega)~=~-\frac{\alpha^2
(N-1)\omega_p^2 \omega^2 g_0(\omega)}{
\omega^2+\omega_0^2+\alpha^2 (N-1)\omega_p^2 \omega^2
g_0(\omega)}f(\omega)~,
\end{equation}
where $g_0(\omega)$ and $f(\omega)$ are the  Fourier-transform of
$G_0(\tau)$ and $f(\tau)$, respectively. The bounce solution
$f(\tau)$ localized on the junction $l$ is determined
self-consistently from the equation:
\begin{eqnarray} \label{Centralsolution}
\frac{1}{\omega_p^2}\ddot{f}+ \alpha^2 \int_0^\infty d
\tau_1G_1(\tau-\tau_1)
\ddot{f}(\tau_1)-(1-j)+\frac{f^2}{2}=0, ~~~~\nonumber \\
G_1(\tau)=\int_{-\infty}^{\infty}\frac{d\omega}{2\pi}\frac{(\omega^2+
\omega_0^2)e^{i\omega \tau} }{\omega^2+\omega_0^2+\alpha^2
(N-1)\omega_p^2 \omega^2 g_0(\omega)}.~~~~~
\end{eqnarray}
In the absence of a global coupling the instanton solution is
strongly localized on a particular junction (see schematic in Fig.
1, dashed line), i.e \cite{Ingold}
\begin{equation} \label{Instantonsol0}
\varphi_n(\tau)=f_0(\tau)\delta_{nl}=\frac{3\sqrt{2(1-j)}}{\cosh^2(\omega_0\tau/2)}\delta_{nl}~~.
\end{equation}
Substituting (\ref{Tailsolution}) in the expression
(\ref{Action-eff}) for the effective action $S_{eff}$  and  using
a perturbative approach (similarly to Refs. \cite{EDM,Legett}),
i.e. $f(\tau)~\simeq~f_0(\tau)$, we obtain the MQT escape rate (in
physical units) as
\begin{equation} \label{GammaMQT}
\Gamma_{MQT}~\simeq~\Gamma_{0}\exp{\left[-\frac{72 E_J}{15 \hbar
\omega_p}2^{1/4}(1-j)^{5/4}(1+\chi) \right]}~,
\end{equation}
where
\begin{equation} \label{chi}
\chi=\frac{30\pi \alpha^2 \omega_p^2}{\omega_0^2} \int_0^\infty dx
\frac{x^4(x^2+1)g_0(x)\sinh^{-2}(\pi x)}{x^2+1+\frac{\omega_p^2
\alpha^2}{\omega_0^2}(N-1)x^2g_0(x)}~,
\end{equation}
where
\begin{equation}
g_0(x)=1/[x^2+(\omega_R/\omega_0)^2]~.
\end{equation}
Here, the parameter $\Gamma_0$ is just proportional to $N$. A
parameter $\chi$ having a positive value, characterizes a
suppression of the MQT due to the presence of the charge
interaction between Josephson junctions of the array and an
external shunting impedance. For short Josephson junction array
($N~\simeq~1$), such a MQT suppression can be rather large for
moderate values of $\alpha$. However, as we turn to large
Josephson junction arrays ($N \gg 1$) a standard MQT behaviour is
recovered. Quantitatively an enhancement of MQT depends strongly
on parameters $\alpha$ and $\omega_R$. The expression (\ref{chi})
can be simplified in two  limits: $\omega_R \gg \omega_0$
("capacitative impedance") and $\omega_R \ll \omega_0$ ("inductive
impedance") as
\begin{equation} \label{chi-limits}
\chi = 5\alpha^2 \omega_p^2 \left\{ \begin{array}{ll}
\frac{1}{\omega_R^2+ \alpha^2 \omega_p^2(N-1)} & \mbox{$\omega_R \gg \omega_0$};\\
      \frac{1}{\omega_0^2+ \alpha^2 \omega_p^2(N-1)}  & \mbox{$\omega_R \ll \omega_0$}.\end{array} \right.
\end{equation}

Typical dependencies of the MQT escape rate $\Gamma_{MQT}$ on the
dc bias current $I$ for various values of $N$ are presented in
Fig. 3. One can see a giant enhancement of the MQT escape rate as
we turn from short to long Josephson junction arrays. This
enhancement results from a decrease of the slope of the bias
current dependence escape rate. Comparing our theoretical
predictions with the experimental curves published in Ref.
\cite{HTSC-QM2} (see Fig. 5 in Ref. \cite{HTSC-QM2}) we find a
good agreement for both the crossover temperature $T^*$ and the
dependence of $\Gamma_{MQT}(I)$ for the inductive type of a
shunting impedance. Therefore, in order to choose between two
models, i.e. a nearest-neighbor intrinsic charge interaction or
external global charge coupling, one needs additional independent
measurements of Debye screening length \cite{Tachicki} or to tune
the MQT by variation of $Z$.

\begin{figure}
\includegraphics[width=2.5in,angle=0]{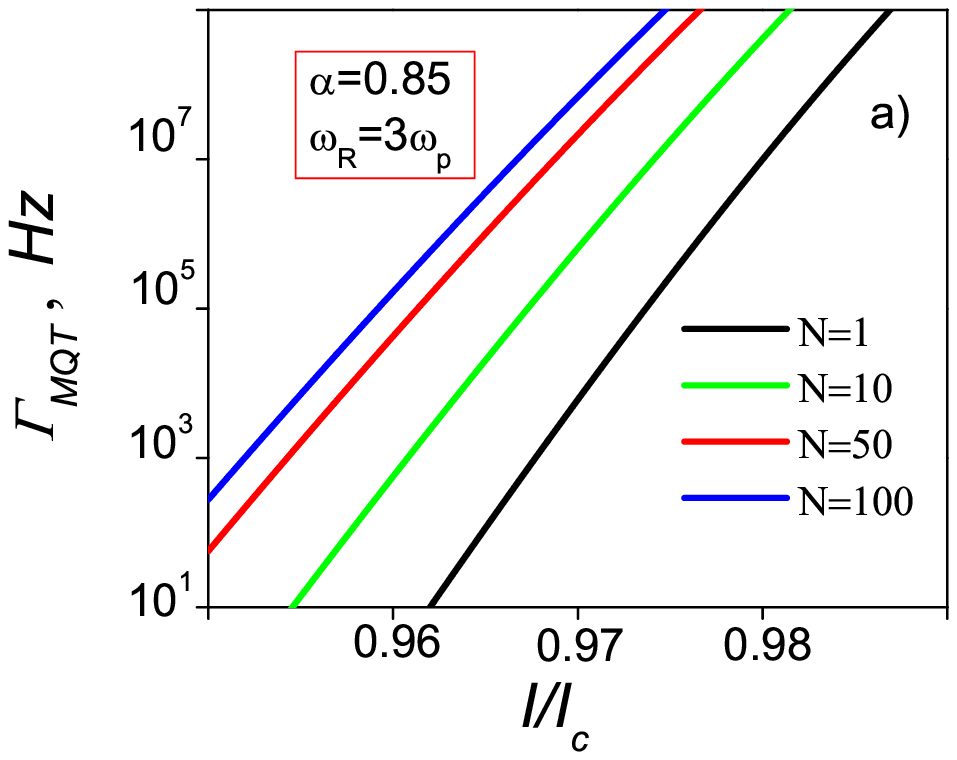}
\includegraphics[width=2.5in,angle=0]{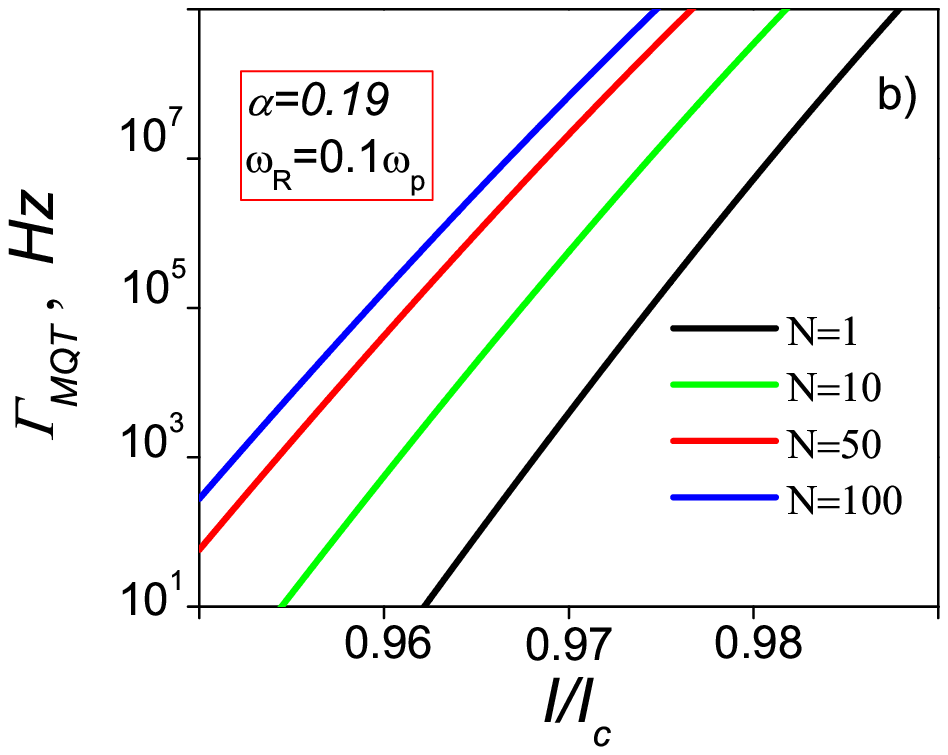}
\caption{The dependence of the MQT escape rate $\Gamma_{MQT}$ on
the dc bias current $I$ for various values of $N=1, 50, 100$. Both
cases of capacitative  (a) and inductive (b) impedance for
particular sets of parameters are shown. }\label{MQTratedep}
\end{figure}

In conclusion we have shown that the dissipative (decoherence)
effects can be strongly suppressed in long ($N>>1$) )Josephson
junction series arrays with a global charge interaction. The both
dissipation and global charge interaction can be introduced
through an external shunting impedance. This effect manifests
itself as a giant enhancement of the MQT escape rate for the
switching from the superconductive state to a resistive one (see
Fig. 3). A giant MQT enhancement is explained through an
excitation of spatial-temporal charge instanton  distributed over
a whole array.

I would like to thank P.M\"uller, and A. V.
Ustinov for useful discussions. I acknowledge the
financial support by SFB 491.

\end{document}